# Magnonic Full Adder Based on 2D Chiral Magnonic Resonators

K. G. Fripp[1], Y. Wang[2], O. Kyriienko[2], A. V. Shytov[1], and V. V. Kruglyak[1]

[1]University of Exeter, Stocker Road, Exeter, EX4 4QL, United Kingdom

[2]University of Sheffield, Western Bank, Sheffield, S10 2TN, United Kingdom

We use micromagnetic simulations to demonstrate how machine learning can be applied to arrays of chiral magnonic resonators to build a magnonic full adder. The chiral magnonic resonators have form of nano-sized permalloy disks that nonlinearly scatter spin waves propagating in a YIG waveguide. The spin waves are injected from multiple outputs, and the dynamic stray magnetic field of the scattered spin waves is sampled in multiple locations to form several physical output signals. These signals are weighted and combined, either linearly or nonlinearly, to satisfy the logic output of a full adder. The process is known as training and forms the device's "output layer". The full adder's performance is evaluated in terms of robustness to input and output noise for a given number of physical output signals and the form of the output layer. When the output layer is linear, as few as three physical signals per logical output are sufficient. When a multilayer perceptron neural network forms the output layer, the number of required output signals is reduced to one, and a nearly perfect classification accuracy is achieved when appropriate preprocessing and augmentation strategies are used.

As the demand for computational power grows, so does the complexity of logic circuits within CMOS devices. This exposes an obvious bottleneck in their architecture – the non-locality of computation and memory, which leads to a large penalty in performance and energy consumption. To alleviate this limitation, a number of alternative computing paradigms, known as "beyond von Neumann" architectures, have been proposed. Among others, neuromorphic computing devices imitate biological brains using synthetic analogues of neurons and synaptic connections within artificial neural networks (ANNs) [1]. Albeit possible to implement using conventional CMOS-based computers, ANNs networks will excel even more strongly on bespoke hardware platforms that bring both the memory and computation into each other's vicinity.

A suitable hardware platform for neuromorphic computing could be offered by magnonics – the science of magnons, spin waves and their applications [2]. The low energy nature of magnons (quanta of spin waves [3,4]) reduces the power associated with the information they carry. Progress in fabrication of low-damping magnonic media, such as thin films of yttrium-iron garnet (YIG) [5,6], ensures that only a small fraction of the power is dissipated as spin waves propagate. The ohmic losses and joule heating are then confined to devices' electrical leads and could be reduced further when voltages eventually replace currents [7,8] at points of signal interconversion and bias application. Miniaturisation of magnonic devices is enabled by short wavelengths of spin waves in the GHz frequency range [3,4], while their inherently strong nonlinearity may be further enhanced in structures that concentrate resonantly the energy of incident spin waves [9,10].

Recent research into magnonic and more generally magnetic neuromorphic computing has seen multiple demonstrations of reservoir computers of varied construction and operation [11-19]. Reservoir computing has the attractive property of training (weighting) only the output layer of an ANN or indeed outputs of any suitable nonlinear "reservoir". In the design of magnonic reservoirs, there is flexibility in the operation of the physical 'reservoir' and the manifestation of its output. The latter could be either directly derived from a linear weighted superposition of the reservoir's output neurons or be fed into another neural network.

A magnonic logic gate performs a Boolean operation on individual or multiple binary spin-wave inputs to produce a single binary output. By combining multiple magnonic logic gates, combinational logic circuits may be created where the binary output or outputs satisfy a more complicated logical function in an analogous way to CMOS technology. Several concepts of magnonic logic circuits have been developed [20-27]. Some of these use the inverse design approach [26,27], which amounts to training an ANN's internal layers. The approach has successfully led to demonstration of magnonic NOT, OR, NOR, AND, NAND, and half-adder gates reprogrammed using a 7×7 array of antennas each fed by ~100 mA bias current to deliver the required graded magnonic landscapes [26]. It is conceivable that that smaller currents could be used when the device is optimised. It is even easier to conceive that a reservoir computer using a similar array of antennas to read out signals due to spin waves scattered by a passive nonlinear reservoir would prove yet more energy efficient.

Of particular interest for creating nonlinear magnonic reservoirs are arrays of subwavelength spin-wave scatterers that each possesses a neuron-like nonlinear response, e.g. the two-dimensional (2D) chiral magnonic resonator from [28]. Specifically, the resonator represents a nanoscale metallic disk hosting a low-frequency edge mode [29]. The mode's reduced frequency and volume match the requirements for a wide-angle resonant scattering of spin waves propagating in a low loss YIG film beneath. The incident spin waves generate a nonlinear shift of the edge mode frequency, which in turn manifests in the scattering pattern and produces an effective activation-like behaviour [28]. A network of such neurons (spatially separated to ensure they interact only via scattered spin waves) may then be expected to provide a suitably complex, nonlinear interference pattern required for an efficient reservoir computation. The disks in the inner layers of the network retain all the useful traits of chiral magnonic resonators [30-33], among which chirality, magnetic re-programmability, and electric tuneability promise reach opportunities for inverse design of the network's nonlinear response.

| Input (ABC) | 000 | 100 | 010 | 110 | 001 | 101 | 011 | 111 |
|---|---|---|---|---|---|---|---|---|
| $C_{\text{out}}$ | 0 | 0 | 0 | 1 | 0 | 1 | 1 | 1 |
| Sum | 0 | 1 | 1 | 0 | 1 | 0 | 0 | 1 |

Table 1. Truth table of a full adder. The different combinations of inputs A, B and C are related to the corresponding binary outputs $C_{\text{out}}$ and Sum.

In this paper, we combine the neuron-like behaviour of 2D chiral magnonic resonators from [28] with the reservoir computing and machine learning techniques to construct a magnonic analogue of a binary 1-bit full adder (Table 1) – the core constituent of larger logic circuits such as arithmetic logic units (ALUs), where additions of multiple-bit binary numbers are performed. The functionality of the magnonic full adder is demonstrated using micromagnetic simulations run with MuMax3 [34]. As shown in Fig.1(a), the device consists of (i) three spin-wave inputs (two 1-bit binary numbers and a carry bit) in the input layer, (ii) six physical outputs (combined into two logical outputs – a sum bit and a carry out bit), and (iii) a nonlinear reservoir formed by a 2×4 array of 2D chiral magnonic resonators that scatter spin waves propagating in a YIG waveguide. The physical inputs are modelled as a local time-dependent magnetic field assumed to be created by inductive antennas or otherwise (e.g. an array of chiral magnonic resonators or spintronic oscillators). The physical output signals are given by the stray magnetic field due to the scattered spin waves. The field is assumed to be read out by inductive antennas or otherwise (e.g. chiral magnonic resonators serving as free layers of magnetoresisitive sensors). The physical signals are combined (linearly or non-linearly) with weights chosen to satisfy the truth table for a full adder (Table 1) or another logic circuit targeted. In the reservoir computing's terms, this process represents training of the computer's output layer.

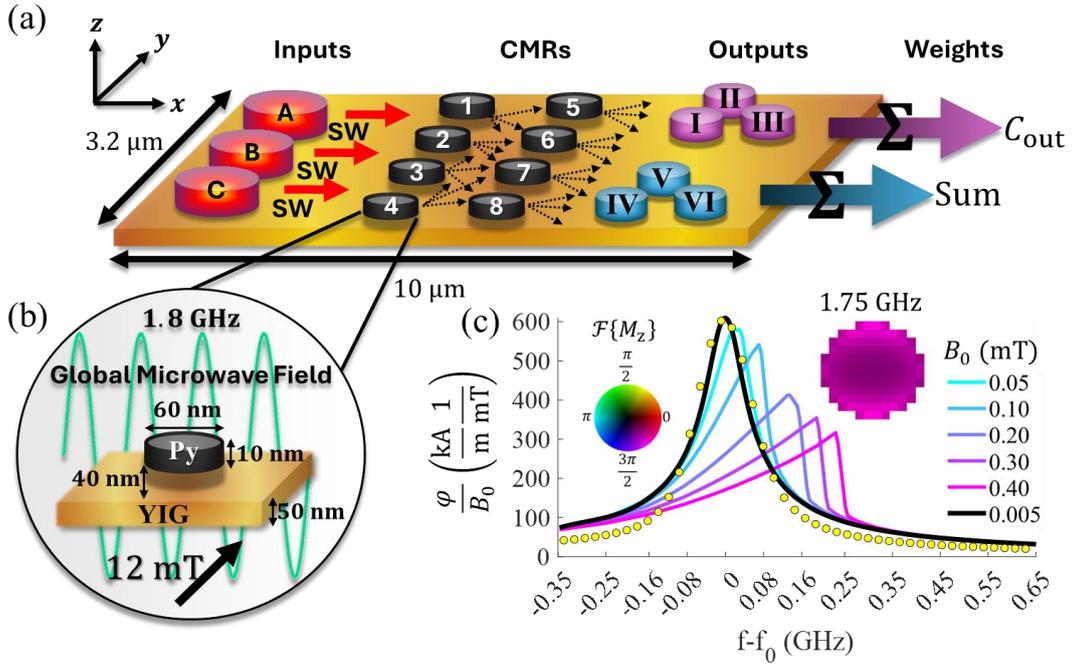

Figure 1. (a) Spin waves injected via the sources A, B and C are scattered nonlinearly from the array of 2D chiral magnonic resonators 1-8. The stray field signals are read-out from the physical outputs I-VI and linearly weighted to form the logical outputs for a full adder (Table 1). (b) Each 2D resonator is a Permalloy disk of 60 nm diameter and 10 nm thickness, situated 40 nm above a 50 nm thick YIG waveguide of 3.2 μm width and 10.24 μm length. The sample is biased by an in-plane magnetic field of 12 mT applied parallel to the $y$ axis and a global microwave field with frequency of 1.8 GHz and magnitude of 0.09 mT polarised along the $x$ axis. (c) The amplitude $\varphi$ of the linear (black curve) and nonlinear (coloured curves) response of the edge mode resonance of an isolated nano disk are shown for different excitation amplitudes $B_0$. In each case, the bias field is tuned to ensure that the resonance frequency is equal $f_0 = 1.75$ GHz, i.e. that of the disk with the waveguide beneath. For a reference, the yellow circles show the spectrum acquired using a broadband excitation of the disk with the waveguide beneath, with the trace re-scaled so that its maximum matches that in the linear response without the waveguide (black curve). The inset shows the resonance mode's spatial profile.

The device's performance (error incurred) and robustness to input / output 'noise' are quantified in terms of the number of combined physical output signals and the complexity of the output layer required to obtain the target logical output. We find that three physical signals per logical output are sufficient to create a viable full adder when the output layer is linear, with a further reduction in the error as the number of physical outputs increases. The configuration of three signals per logical output is robust to perturbation of the input amplitudes of the order of 10%. To go beyond linear weighting, we use a multilayer perceptron neural network [35] with the output layer trained for signal classification from just a single physical output signal. The network's output layer has 16 neurons, one half of which encodes the output of the full adder and the other – the error correction. When combined with appropriate preprocessing and augmentation strategies [36,37], the network demonstrates a nearly perfect classification of the physical output signal, with a strong resilience to noise. We compare the linear and nonlinear output layers in terms of their performance and feasibility towards realisation of a magnonic full adder.

Figure 1 illustrates the device geometry and the nonlinear properties of the constituent 2D chiral magnonic resonators, with additional details reported in [28]. In the micromagnetic simulations, the sample is discretised into cuboidal cells with edge length of 5×5×10 nm$^3$ on a regular mesh of 2048×2048×10 ($x \times y \times z$) size. The 'edge-smoothing' of 4$^{th}$ order [34] is adopted to correct the discretisation error at the resonators' edges. The mode spectrum and character of nanoscale magnetic elements are sensitive both to the edge uniformity experimentally [38] and the discretisation numerically [39], which should be considered when using our results to build practical devices. Figure 1(c) shows the response of the edge mode resonance of an isolated disk to excitation by a uniform circularly polarised harmonic magnetic field

$$B_{\text{ex}}(t) = B_0\big(\hat{x}\sin(2\pi f t) + \hat{z}\cos(2\pi f t)\big), \tag{1}$$

where $B_0$ and $f$ are the field's amplitude and frequency, respectively. The response is strongly nonlinear, with the resonance visibly shifting towards higher frequencies already at modest excitation amplitudes ($B_0 \gtrsim 0.1$ mT). The excitation field's circular polarisation mimics that of the dynamic stray field of propagating plane spin waves. So, the disk's response to this field provides a reference when considering the nonlinear spin wave scattering patterns later.

The disk array is situated above a thin YIG waveguide. A uniform bias field of 12 mT aligns the magnetisation in the disks and the waveguide parallel to the $y$ axis (Fig. 1(b)). We introduce a parabolic-shaped Gilbert damping profile at either end of the waveguide in the $x$ direction to prevent re-circulation and reflection of spin waves from the waveguide ends. This contrasts with the $y$ direction, in which no absorbing or periodic boundary conditions are applied. Instead, the finite width of the waveguide helps re-direct spin waves towards the disk array and device's outputs.

In each input of the magnonic full adder, the spin waves are excited by a harmonic circularly polarised field with a constant frequency of 1.8 GHz and a two-dimensional, Gaussian spatial profile given by

$$I_i(x, y) = A_i \cdot \exp\left(-\frac{(x-x_i)^2 + (y-y_i)^2}{2\sigma^2}\right), \quad (2)$$

where subscript $i$ is 'A', 'B', or 'C', $A_i$ is the field amplitude, $x_A = x_B = x_C = -2\ \mu m$, $y_A = 0.65\ \mu m$, $y_B = 0\ \mu m$, and $y_C = -0.65\ \mu m$. The low-state ('0') and high state ('1') amplitudes are $0.18\ mT$ and $0.45\ mT$, respectively. $\sigma$ corresponds to a full width at half maximum (FWHM) of $350\ nm$, which facilitates spin wave emission into a wide angle. In addition to the local spin wave inputs, we excite the system by a global uniform microwave field of $0.09\ mT$ amplitude, polarised along the $x$ direction. This field couples directly to the disks (even without any input present) with each of them acting as a low-amplitude spin wave source. Also, the same field leads to spin wave generation from the waveguide's lateral edges (parallel to the $x$ axis) through a mechanism similar to that described e.g. in [40]. We have found that this additional power injection has a positive effect on the device performance.

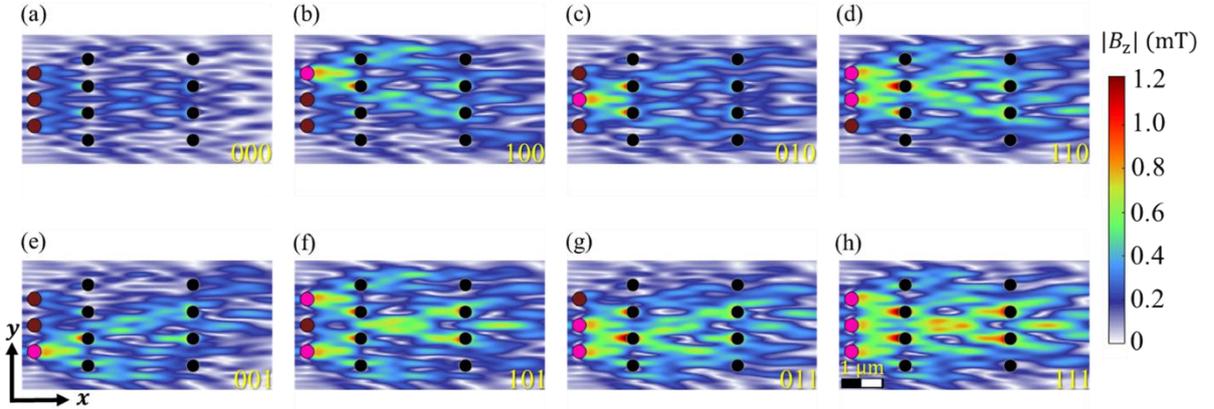

Figure 2. Panels (a) to (h) show patterns of the stray magnetic field, $|B_z|$, created by spin waves propagating and scattered in the waveguide at a height of $40\ nm$ above its surface, for each of the possible input combinations. In each panel, the high ("1") and low ("0") excitation states of the inputs A, B, and C (top to bottom) are indicated by the pink and dark red circles, respectively. The positions of the resonators are indicated by the black circles (not to scale).

Figure 2 shows the patterns of the absolute value of the Fourier amplitude of the $z$ component of the stray magnetic field created by spin waves in the device, for each of its input combinations from Table 1. The '000' input combination means that all three input field amplitudes are set to the 'low' state ($0.18\ mT$). The corresponding nonuniform pattern shown in Fig 2(a) does not significantly differ from a linear one. The superposition of spin waves launched from the three sources leads to several regions of constructive interference coinciding (by design) with the positions of the resonators both in the first and second columns of the array. The positions are located at intersections of the caustic beams. The latter are formed due to the anisotropy of the spin wave dispersion in the in-plane-magnetised thin-film waveguide [41] and because both the inputs and the resonators serve as nearly perfect point sources of primary and secondary waves, respectively. The wave pattern changes for the '111' input combination (Fig. 2(h)), which has the same symmetry while all three inputs are set to the 'high' state ($0.45 \text{ mT}$). The interference maxima have moved relative to their '000' positions, suggesting a nonlinear, position-specific dependence on the excitation strength.

Let us now consider the interference patterns formed when just one of the inputs is in the 'high' state ('1'), i.e. is 'ON'. Shown in Fig. 2(b), (c) and (e), the patterns are dominated by the two caustic beams with greater amplitude emanating from the ON source. When an off-centre input is ON (Fig. 2(b), (e)), one of its caustic beams hits the boundary of the waveguide, is reflected from it, and is incident upon a resonator in the second column of the array. The other caustic beam is incident upon a resonator in the first column, with an increased fraction of its energy transmitted through (as compared to the small amplitude case) due to 'activation' caused by the nonlinear frequency shift of the edge mode resonance (Fig.1(c)). When the centre input is ON (Fig. 2(c)), each of its two caustic beams intersects and get transmitted through a resonator in the first column. When two inputs are ON (Fig. 2 (d), (f), and (g)), additional resonators are activated, and the interference pattern on the output side of the array is modified again.

Figure 2 demonstrates that each input combination results in a distinct and complex interference pattern arising from the nonlinear scattering of spin waves by the array of disk resonators. We can sample (e.g. inductively) the dynamic stray magnetic field at multiple locations in the patterns and sum the signals (with weighting, which can either be built into the antenna configuration, or done externally) to create the logical output for a full adder (Table 1). Notably, even the 'high' input state at which the disks experience a strong nonlinear frequency shift of their edge mode resonances (Fig. 1(c)) are quite low in absolute terms ($0.45 \text{ mT}$), promising enhanced energy efficiency of the device. Furthermore, the patterns are highly non-uniform in amplitude and phase (not shown) even at the lowest excitation strengths. This is advantageous, as firstly, we want an output when the inputs are all 'OFF', and secondly, the complex spatial dependence offers an extensive selection of nonlinear functions (at different sampling locations) available to achieve the targeted functionality of the device. The strong nonlinearity underpins the effectiveness of training the linear output layer (finding the optimal weighting in a linear superposition of the functions), i.e. the computational power occurs due to the nonlinearity in the magnonic system. Alternatively, we can extract the signal just from a single location and analyse it using a nonlinear weighting in a neural network to derive the sought logical output. We refer to the latter approach as a nonlinear output layer.

The approach for the linear output layer is as follows. For each of the eight input combinations (Table 1), the complex-valued Fourier amplitude calculated from the $z$ component of the spin waves' dynamic stray field (Fig. 2) is convolved with a Gaussian kernel with FWHM of $190$ nm. This mimics the effect of a finite size of detecting antennae. $2N$ physical sampling points (where the antennae are assumed to be located) are selected, $N$ each for the $C_\text{out}$ or Sum logical outputs for the full adder (Table 1). For each sampling point, the complex field values obtained for the eight input combinations are used to form a sampling column vector of length 8. Then, separately for $C_\text{out}$ and Sum, (i) these are combined into a rectangular, complex-valued $8 \times N$ matrix $\boldsymbol{A}$, whose $N$ columns represent sampling vectors for the $N$ sampling points and whose 8 rows correspond to the different input combinations, (ii) a target output vector $\boldsymbol{b}$ (of length 8) with appropriate thresholds for logical 'low' and 'high' values is defined, and (iii) a set of linear equations in the unknown weights vector $\boldsymbol{W}$ (of length $N$) is constructed as

$$\boldsymbol{A} \cdot \boldsymbol{W} = \boldsymbol{b}. \qquad (2)$$

Equation (2) must be augmented by realistic assumptions about the physical structure of the device. Here, we assume that the $C_\text{out}$ or Sum outputs are obtained using an envelope detector to rectify the signal from an inductive antenna formed by connecting in series individual Ω-shaped detecting antennae with their size controlling the weight of contributions from the individual sampling points. As the electromagnetic wavelength is much longer than the device size, we restrict our search to real-valued weights $\boldsymbol{W}$, while the use of an envelope detection renders the phase of $\boldsymbol{b}$ irrelevant. Hence, we replace Eq. (2) by a nonlinear equation

$$\boldsymbol{r} = \text{Re}[\boldsymbol{A} \cdot \boldsymbol{W}] \odot \text{Re}[\boldsymbol{A} \cdot \boldsymbol{W}] + \text{Im}[\boldsymbol{A} \cdot \boldsymbol{W}] \odot \text{Im}[\boldsymbol{A} \cdot \boldsymbol{W}] - \boldsymbol{b} \odot \boldsymbol{b}, \qquad (3)$$

for real-valued $\boldsymbol{W}$, where $\odot$ is the Hadamard (component-wise) product. We solve Eq. (3) repeatedly for different locations of $N$ sampling points, to minimise globally the error defined as

$$\boldsymbol{\epsilon} = (\boldsymbol{r} \odot \boldsymbol{r})^{1/4}, \qquad (4)$$

where $\boldsymbol{\epsilon}$ is a vector of length 8 of errors incurred from all input combinations. We enforce that for the optimal solution of Eq. 3 for $N$ sampling points, the largest magnitude component of $\boldsymbol{\epsilon}$ globally is minimised, $\|\boldsymbol{\epsilon}\|_\infty$, as opposed to the L2 norm (average magnitude) of the component errors, $\|\boldsymbol{\epsilon}\|_2$. In other words, we prioritise the overall functional accuracy of the device over its precision for individual input combinations. Finding the best combination of $N$ sampling points is a combinatorial optimisation problem solvable by a variety of methods. To tackle the factorial complexity, we adopt different approaches for smaller $N$ ($\leq 5$) and larger $N$ ($> 5$). These are referred to as a parallel grid search and an iterative-thresholded search, respectively, and are described in the Supplementary material.

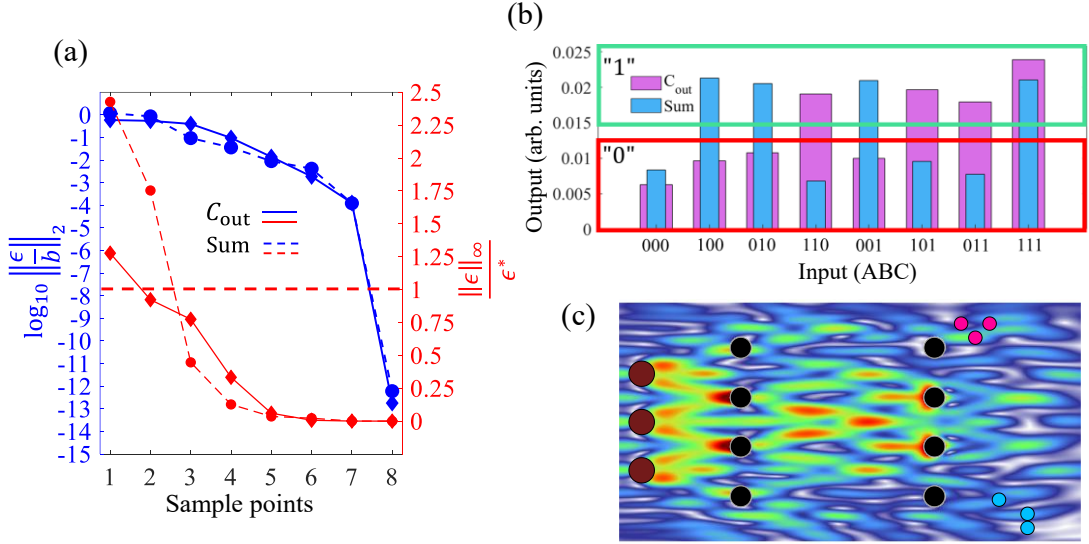

Figure 3. (a) The L2 norm of the relative error $\epsilon/b$ ($\|\epsilon/b\|_2$, left vertical axis) and the normalised maximum $\epsilon$ for different input combinations ($\|\epsilon/\epsilon^*\|_\infty$, right vertical axis) are shown as a function of the number of sampling points. The red dashed line corresponds to $\|\epsilon\|_\infty = \epsilon^* = 0.004$ (arb. units), with values $\|\epsilon/\epsilon^*\|_\infty < 1$ indicating that the full adder's outputs are valid for all input combinations. (b) A bar chart of the output obtained using three sampling points per logical output, for $C_{\text{out}}$ (blue) and Sum (pink). The green and red regions indicate the logical ranges adopted for the low and high state outputs. (c) The optimal positions of the three sampling points relative to those of the inputs and the resonator array are shown for the $C_{\text{out}}$ (blue) and Sum (pink) outputs. The sampling positions are overlaid onto the stray magnetic field for the "000" input configuration.

Figure 3 presents our optimal solutions obtained for different numbers of the sampling points. Figure 3 (a) shows the dependence of the L2 norm of the relative error, $\|\epsilon/b\|_2$, and the normalised maximum error for different input combinations $\|\epsilon/\epsilon^*\|_\infty$ on the number of sampling points. The norm of the error decreases with the number of sampling points, as expected. For 7 and 8 sampling points, $\|\epsilon/b\|_2$ is effectively 0 (machine precision). We, however, desire as few sampling points as possible to reduce the complexity of the linear output layer. Both for $C_{\text{out}}$ and Sum outputs, we find three sampling points to be the fewest possible whilst all outputs remain valid (Fig. 3(b)). There is a degree of symmetry here. Indeed, the device takes 3 bits as the input, and so, it may be expected that 3 bits (three sampling points) would be required to correctly define each output. Fig. 3(b) shows the result obtained using three sampling points per output, with the locations of the sampling points / detecting antennae shown in Fig. 3(c). We see that the height of the bars corresponds approximately to the middle of the range assigned for the high and low output states.

Let us now evaluate robustness of the configuration with the three sampling points per output (Fig. 3(b)) with respect to noise in the inputs, i.e. random variation of the field amplitudes in the sources. Specifically, we perform additional micromagnetic simulations (5 per input combination) in which the input field amplitudes are varied quasi-randomly by about $\pm 0.018$ mT (±10% of the 'low' input amplitude $0.18$ mT). The results are analysed as before and compared to those shown in Fig. 3 (b). The outputs for all input combinations remain within the specified logic ranges, while the average error has increased by about 40% relative to those shown in Fig. 3(a).

Let us now consider a nonlinear alternative to the case of the linear weighting considered above, when the computation results from the nonlinearity of the magnonic system. We employ a neural network as the nonlinear output layer of the device and show how it can classify the signal derived from just a single sampling point. The trade-off is the increased complexity of the output layer, now implemented in silico as described in the Supplementary material. The output layer is implemented using a multilayer perceptron (MLP) neural network with configuration shown in Fig. 4(a). The signal is assumed to be measured by a single detecting antenna (shown as "$C$" in Fig. 4(a)) and is split into its real and imaginary parts using e.g. an IQ mixer circuit. The resulting two signals are fed into the MLP network. The network comprises two hidden layers with 64 and 32 neurons, respectively, each using a ReLU (Rectified Linear Unit) activation function. The output layer contains 16 neurons with softmax activation: eight outputs correspond to the eight full-adder input classes, while the remaining eight outputs that estimate the classification error. This nonlinear readout allows the device to recover the required logic outputs from a single physical signal, at the cost of additional (electronic) post-processing complexity.

Let us now evaluate robustness of the configuration with a MLP network output layer to noise. Figure 4(b) presents our results of the averaged classification accuracy of the MLP output layer for noisy signals $C_\text{noisy}$, which are defined as

$$C_\text{noisy} = C + r_\text{noise} \times |\bar{C}|\big(1 - 2R_1 + i(1 - 2R_2)\big) , \qquad (5)$$

where $C$ is the noise-free signal (convolved with a Gaussian kernel with FWHM of $190$ nm, as before), $|\bar{C}|$ is the maximum value of the intensity map within the region of the detector, $r_\text{noise}$ is a parameter scaling the magnitude of the noise (with $r_\text{noise} = 0$ corresponding to the results of Fig.3 (a)), and $R_1$ and $R_2$ are random numbers generated separately in the range [0,1]. Multiple datasets were generated using different random seeds to acquire different $R_1$ and $R_2$ values. In addition to the noise added in Eq. 5, spatial shifts of the detector were applied ($\pm 5$ nm), where the data used for training were normalised in the range [0,1].

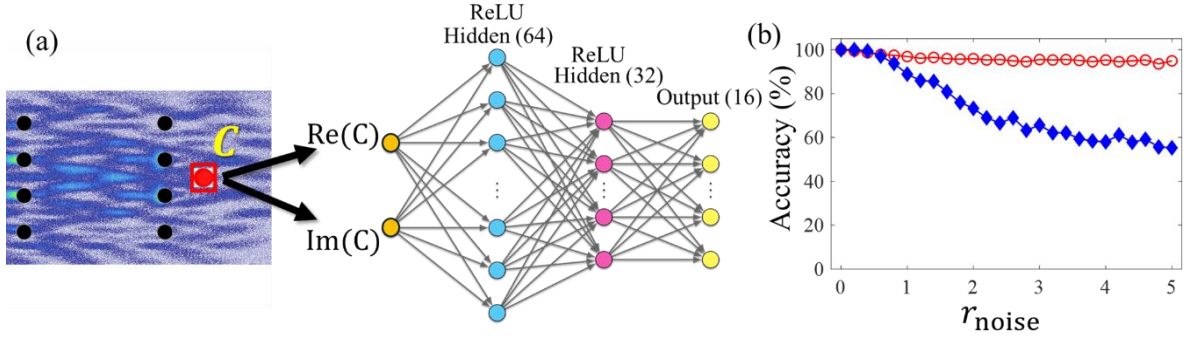

Figure 4. (a) The dynamic field pattern of the stray field from Fig. 2 with complex-valued noise added ($r_{\text{noise}} = 5.0$, see the text) is shown for the input combination "000". The single antenna $C$ (red circle and square) located at $x = 2.6$ μm, $y = 0$ μm detects the spin waves' signal. The real and imaginary parts of the Fourier amplitude of this signal act as inputs to a three-layer MLP network. The first and second hidden layers contain 64 and 32 neurons (units), respectively, where the activation function is the ReLU (Rectified Linear Unit). The output layer produces 16 values (twice that of the full adder), where the additional 8 outputs contain the classification error. (b) Averaged classification accuracy as a function of the noise parameter up to 5.0. The MLP approach (red open circles) maintains near-perfect accuracy across the entire noise range, while the linear approach (blue diamonds, for linear trained at each noise level) degrade significantly for noise ratios above 1.0.

The results of Fig. 4(b) compare two approaches: the MLP network (red open circles), and a linear output layer trained and tested at each $r_{\text{noise}}$ level (blue diamonds). The MLP approach maintains near-perfect accuracy (approximately 95–100%) across the entire $r_{\text{noise}}$ range from 0 to 5. In contrast, the linear approach shows a rapid decline in accuracy: the Linear model drops from near 100% at zero noise to approximately 60% by $r_{\text{noise}}$ of 4.0, remaining around this level till $r_{\text{noise}}$ is equal to 5.0. This demonstrates the superior noise resilience of the MLP approach. Even at $r_{\text{noise}} = 5.0$, the MLP maintains approximately 95% accuracy, indicating that the MLP-based device can effectively handle substantial noise levels. Overall, the MLP network dramatically outperforms the linear output layer demonstrated in Fig. 3 when noise is concerned. This robustness is particularly relevant for practical implementations where thermal fluctuations, fabrication imperfections, and signal noise are unavoidable.

We now draw attention to a comparison of the benefits and drawbacks of using either a linear or nonlinear output layer in a physical implementation of a magnonic full adder. The around 50% accuracy plateau observed for the linear output layer under noise is notable: for an 8-class classification problem, random guessing would yield only 12.5% accuracy, so the linear model retains some discriminative capability but fails to reliably distinguish between all input states, whereas the MLP's nonlinear decision (16-class including error corrections) boundaries remain effective. This suggests that the magnonic reservoir's nonlinear scattering patterns contain information that can only be more effectively extracted through nonlinear readout. On the surface, the performance obtained using the nonlinear output layer suggests it is ideal. Yet, its extra complexity will require a small-scale, likely CMOS device, to perform the operations of the MLP network. In contrast, the signal weighting in the linear output layer is likely to require a simpler (and so, less power hungry) device or could even be performed passively via linear interference of the sampled signals. In addition, the electrical circuit behind an envelope detector used with the linear output is much simpler that required for IQ detection needed for the nonlinear output layer to function. The interference and weighting could be controlled by the construction of and connections between the individual detecting antennae themselves, with several options available (omega-shaped antennas [19], magneto-resistive sensors [42], or other chiral magnonic resonators [30]). The use of a small external circuit (either with a linear or nonlinear output layer) provides a degree of flexibility, in case parameters of the device change (inputs, truth table, etc). Indeed, the signals read out at the sampling locations could be used to acquire new weights, as the computational effort to solve the nonlinear equations is low. There however is no guarantee that the positions selected for the fixed sampling points are the optimal configuration for such parameter changes. The nonlinear output layer would be more versatile in such cases, as it can be more easily re-trained to learn new features in the data.

In summary, using micromagnetic simulations, we demonstrate the functionality of a 1-bit magnonic full adder with performance derived from the ability of an array of nano-disk chiral magnonic resonators to scatter nonlinearly incident spin waves in a YIG magnonic waveguide. We describe how the signals derived from the stray magnetic field due to the scattered spin waves and read out by just three detecting antennae could be weighed and linearly combined to satisfy the logical truth table for a full adder. Alternatively, the function can be achieved feeding the signal derived from a single detecting antenna to an external MLP network. The robustness of the magnonic full adders to input/output 'noise' is quantified. We find that the configuration with three sampling points per logical output and a linear output layer is robust to perturbation of the input amplitudes of about $10\%$. The nonlinear MLP output layer outperforms the linear layer: the former maintains a near 100% accuracy in the noise range in which the latter's accuracy degrades to almost 50%. The results presented here can be extended to more complex multiple input / multiple output devices [43].

The research leading to these results has received funding from the UK Research and Innovation (UKRI) under the UK government's Horizon Europe funding guarantee (Grant No. 10039217) as part of the Horizon Europe (HORIZON-CL4-2021-DIGITAL-EMERGING-01) under Grant Agreement No. 101070347. Yet, views and opinions expressed are those of the authors only and do not necessarily reflect those of the EU, and the EU cannot be held responsible for them.

[1]     Y. LeCun, Y. Bengio and G. Hinton, *Deep Learning*, Nature **521**, 436 (2015).

[2]     *TRTM Dictionary of Magnonics 2025* (TRTM Community, June 2025); https://trtm2025.sciencesconf.org/resource/page/id/22, accessed on 6 January 2026.

[3]     A. G. Gurevich and G. A. Melkov, *Magnetization Oscillations and Waves* (Chemical Rubber Corporation, New York, 1996).

[4]     A. Prabhakar and D. D. Stancil, *Spin Waves* (Springer New York, 2009).

[5]     G. Schmidt, C. Hauser, P. Trempler, M. Paleschke, and E. T. Papaioannou, *Ultra Thin Films of Yttrium Iron Garnet with Very Low Damping: A Review*, Phys. Status Solidi B **257**, 1900644 (2020).

[6]     A. R. Will-Cole, J. L. Hart, V. Lauter, A. Grutter, C. Dubs, M. Lindner, T. Reimann, N. R. Valdez, C. J. Pearce, T. C. Monson, J. J. Cha, D. Heiman, and N. X. Sun, *Negligible magnetic losses at low temperatures in liquid phase epitaxy grown $Y_3Fe_5O_{12}$ films*, Phys. Rev. Mater. **7**, 054411 (2023).

[7]     Y.-J. Chen, H. K. Lee, R. Verba, J. A. Katine, I. Barsukov, V. Tiberkevich, J. Q. Xiao, A. N. Slavin, and I. N. Krivorotov, *Parametric Resonance of Magnetization Excited by Electric Field*, NanoLett. **17**, 572 (2017).

[8]     B. Rana, *Role of voltage-controlled magnetic anisotropy in the recent development of magnonics and spintronics*, J. Appl. Phys. **136**, 150701 (2024).

[9]     K. G. Fripp, Y. Au, A. V. Shytov, and V. V. Kruglyak, *Nonlinear chiral magnonic resonators: Toward magnonic neurons*, Appl. Phys. Lett. **122**, 172403 (2023).

[10]    A. Lutsenko, K. G. Fripp, L. Flajšman, A. V. Shytov, V. V. Kruglyak, and S. van Dijken, *Nonlinear dynamics in magnonic Fabry-Pérot resonators: Low-power neuron-like activation and transmission suppression*, arXiv:2602.10650 (2026).

[11]    J. Torrejon, M. Riou, F. Abreu Araujo, S. Tsunegi, G. Khalsa, D. Querlioz, P. Bortolotti, V. Cros, K. Yakushiji, A. Fukushima, H. Kubota, S. Yuasa, M. D. Stiles, and J. Grollier, *Neuromorphic computing with nanoscale spintronic oscillators*, Nature **547**, 428 (2017).

[12]    R. Nakane, G. Tanaka, and A. Hirose, *Reservoir Computing with Spin Waves Excited in a Garnet Film*, IEEE Access **6**, 4462 (2018).

[13] S. Watt, M. Kostylev, A. B. Ustinov and B. A. Kalinikos, *Implementing a Magnonic Reservoir Computer Model Based on Time-Delay Multiplexing*, Phys. Rev. Appl. **15**, 064060 (2021).

[14] Á. Papp, W. Porod and G. Csaba, *Nanoscale neural network using non-linear spin-wave interference,* Nat. Commun. **12**, 6422 (2021).

[15] R. V. Ababei, M. O. A. Ellis, I. T. Vidamour, D. S. Devadasan, D. A. Allwood, E. Vasilaki, and T. J. Hayward, *Neuromorphic computation with a single magnetic domain wall*, Sci. Rep. **11**, 15587 (2021).

[16] J. C. Gartside, K. D. Stenning, A. Vanstone, H. H. Holder, D. M. Arroo, T. Dion, F. Caravelli, H. Kurebayashi, and W. R. Branford, *Reconfigurable training and reservoir computing in an artificial spin-vortex ice via spin-wave fingerprinting*, Nature Nanotechnol. **17**, 460 (2022).

[17] W. Namiki, D. Nishioka, Y. Yamaguchi, T. Tsuchiya, T. Higuchi, and K. Terabe, *Experimental Demonstration of High-Performance Physical Reservoir Computing with Nonlinear Interfered Spin Wave Multidetection*, Adv. Intell. Syst. **5**, 2300228 (2023).

[18] S. Nagase, S. Nezu, and K. Sekiguchi, *Spin-wave reservoir chips with short-term memory for high-speed estimation of external magnetic fields*, Phys. Rev. Appl. **22**, 024072 (2024).

[19] C. Heins, J. V. Kim, L. Körber, J. Fassbender, H. Schultheiss, and K. Schultheiss, *Benchmarking a magnon-scattering reservoir with modal and temporal multiplexing*, Phys. Rev. Appl. **23**, 054087 (2025).

[20] K.-S. Lee and S.-K. Kim, *Conceptual design of spin wave logic gates based on a Mach-Zehnder-type spin wave interferometer for universal logic functions*, J. Appl. Phys. **104**, 053909 (2008).

[21] T. Brächer and P. Pirro, *An analog magnon adder for all-magnonic neurons*, J. Appl. Phys. **21**, 152119 (2018).

[22] A. B. Ustinov, E. Lahderanta, M. Inoue, and B. A. Kalinikos, *Nonlinear Spin-Wave Logic Gates*, IEEE Magn. Lett. **10**, 5508204 (2019).

[23] A. Mahmoud, F. Vanderveken, C. Adelmann, F. Ciubotaru, S. Cotofana and S. Hamdioui, *2-Output Spin Wave Programmable Logic Gate*, IEEE Comp. Soc. Ann. Symp. on VLSI (ISVLSI), Limassol, Cyprus, 60-65 (2020).

[24] F. Schulz, F. Groß, J. Förster, S. Mayr, M. Weigand, E. Goering, J. Gräfe, G. Schütz, and S. Wintz, *Realization of a magnonic analog adder with frequency-division multiplexing*, AIP Adv. **13**, 015115 (2023).

[25] K. O. Nikolaev, D. Raskhodchikov, J. Bensmann, E. Lomonte, L. Jin, R. Schmidt, J. Kern, S. Michaelis de Vasconcellos, R. Bratschitsch, S. O. Demokritov, W. H. P. Pernice, and V. E. Demidov, *Operation of a submicrometer waveguide cross as a spin-wave logic gate*, Appl. Phys. Lett. **123**, 142402 (2023).

[26] N. Zenbaa, F. Majcen, C. Abert, F. Bruckner, N. J. Mauser, T. Schrefl, Q. Wang, D. Suess, and A. V. Chumak, *Realization of inverse-design magnonic logic gates*, Sci. Adv. **11**, 9032 (2025).

[27]    Z. Chen, G. J. Lim, C. C. I. Ang, T. Jin, F. Tan, B. W. H. Cheng, and W. S. Lew, *Voltage-controlled half adder via magnonic inverse design*, Appl. Phys. Lett. **126**, 132406 (2025).

[28]    K. G. Fripp, A. V. Shytov, and V. V. Kruglyak, *Nanoscale Magnonic Neurons*, arXiv:2512.00199 [physics.app-ph] (2025).

[29]    V. V. Kruglyak, A. Barman, R. J. Hicken, J. R. Childress, and J. A. Katine, *Picosecond magnetization dynamics in nanomagnets: Crossover to nonuniform precession*, Phys. Rev. B **71**, 220409 (2005).

[30]    V. V. Kruglyak, *Chiral magnonic resonators: Rediscovering the basic magnetic chirality in magnonics*, Appl. Phys. Lett. **119**, 200502 (2021).

[31]    K. G. Fripp, A. V. Shytov, and V. V. Kruglyak, *Spin-wave control using dark modes in chiral magnonic resonators*, Phys. Rev. B **104**, 054437 (2021).

[32]    Y. Y. Au and K. G. Fripp, *Electric Field Control of Chiral Magnonic Resonators for Spin-Wave Manipulation*, Phys. Rev. Appl. **12**, 034023 (2023).

[33]    C. Cai, Z. Zhang, J. Zou, G. E. W. Bauer, and T. Yu, *Spin-orbit-locked coupling of localized microwaves to magnons*, Phys. Rev. Appl. **22**, 034042 (2024).

[34]    A. Vansteenkiste, J. Leliaert, M. Dvornik, M. Helsen, F. Garcia-Sanchez, and B. van Waeyenberge, *The design and verification of MuMax3*, AIP Adv. **4**, 107133 (2014).

[35]    G. E. Karniadakis, I. G. Kevrekidis, L. Lu, P. Perdikaris, S. Wang, and L. Yang, *Physics-informed machine learning*, Nature Rev. Phys. **3**, 422 (2021).

[36]    D. Marković, A. Mizrahi, D. Querlioz, and J. Grollier, *Physics for neuromorphic computing*, Nature Rev. Phys. **2**, 499 (2020).

[37]    J. Grollier, D. Querlioz, K. Y. Camsari, K. Everschor-Sitte, S. Fukami, and M. D. Stiles, *Neuromorphic spintronics*, Nature Electron. **3**, 360 (2020).

[38]    P. S. Keatley, V. V. Kruglyak, A. Neudert, E. A. Galaktionov, R. J. Hicken, J. R. Childress, and J. A. Katine, *Time-resolved investigation of magnetization dynamics of arrays of nonellipsoidal nanomagnets with nonuniform ground states*, Phys. Rev. B **78**, 214412 (2008).

[39]    R. E. Camley, B. V. McGrath, Y. Khivintsev, Z. Celinski, R. Adam, C. M. Schneider, and M. Grimsditch, *Effect of cell size in calculating frequencies of magnetic modes using micromagnetics: Special role of the uniform mode*, Phys. Rev. B **78**, 024425 (2008).

[40]    F. B. Mushenok, R. Dost, C. S. Davies, D. A. Allwood, B. Inkson, G. Hrkac and V. V. Kruglyak, *Broadband conversion of microwaves into propagating spin waves in patterned magnetic structures*, Appl. Phys. Lett. **111**, 042404 (2017).

[41]    V. Veerakumar and R. E. Camley, *Magnon focusing in thin ferromagnetic films*, Phys. Rev. B **74**, 214401 (2006).

[42]    M. Quinsat, D. Gusakova, J. F. Sierra, J. P. Michel, D. Houssameddine, B. Delaet, M.-C. Cyrille, U. Ebels, B. Dieny, L. D. Buda-Prejbeanu, J. A. Katine, D. Mauri, A. Zeltser, M. Prigent, J.-C. Nallatamby, R. Sommet, *Amplitude and phase noise of magnetic tunnel junction oscillators*, Appl. Phys. Lett. **97**, 182507 (2010).

[43]    M. Gołębiewski, P. Gruszecki, M. Krawczyk, *Self-imaging based programmable spin-wave lookup tables*, Adv. Electron. Mater. **8**, 2200373 (2022).